\documentclass[twocolumn,aps,prb,groupedaddress,superscriptaddress,amsmath,floatfix,amssymb,showpacs,noeprint,english]{revtex4-2}
\usepackage{graphicx}
\usepackage{dcolumn}
\usepackage{dblfloatfix}
\usepackage[caption=false]{subfig} 
\usepackage{bm}
\usepackage[T1]{fontenc}
\usepackage{mathrsfs}
\usepackage{amsbsy}
\usepackage{amstext}
\usepackage{amsmath, amsthm, amsfonts, amssymb, bbold}
\usepackage{setspace}
\usepackage{esint}
\usepackage{xcolor,colortbl}
\usepackage{float}
\usepackage{placeins}
\usepackage{algorithm}
\usepackage{algpseudocode}
\usepackage[colorlinks,citecolor=blue,urlcolor=blue,linkcolor=blue,hypertexnames=true]{hyperref} 
\usepackage{babel}
\usepackage{booktabs}
\usepackage{tikz}

\usepackage{tikz}
\usepackage{quantikz}
\usepackage{threeparttable}
\usepackage{multirow}
\usepackage{float}

\usepackage{url}            
\begin{document}

\title{Warm-Starting PCE for Traveling Salesman Problem}

\author{Rafael~Simões~do~Carmo}
\affiliation{Faculty of Sciences, UNESP - S{\~a}o Paulo State University, 17033-360 Bauru-SP, Brazil}

\author{Renato Gomes dos Reis}
\affiliation{Faculty of Sciences, UNESP - S{\~a}o Paulo State University, 17033-360 Bauru-SP, Brazil}

\author{Samuel Fernando F Silva}
\affiliation{School of Arts, Sciences and Humanities, USP – University of São Paulo, 03828-000 São Paulo-SP, Brazil}

\author{Luiz Gustavo E. Arruda}
\affiliation{QuaTI - Quantum Technology \& Information, 13560-161 São Carlos-SP, Brazil}

\author{Felipe F. Fanchini}
\affiliation{Faculty of Sciences, UNESP - S{\~a}o Paulo State University, 17033-360 Bauru-SP, Brazil}
\affiliation{QuaTI - Quantum Technology \& Information, 13560-161 São Carlos-SP, Brazil}

\date{\today}

\begin{abstract}
Variational quantum algorithms are promising for combinatorial optimization, but their scalability is often limited by qubit-intensive encoding schemes. To overcome this bottleneck, Pauli Correlation Encoding (PCE) has emerged as one of the most promising algorithms in this scenario. The method offers not only a polynomial reduction in qubit count and a suppression of barren plateaus but also demonstrates competitive performance with state-of-the-art methods on Maxcut. In this work, we propose a warm-start PCE, an extension that incorporates a classical bias from the Goemans-Williamson (GW) randomized rounding algorithm into the loss function to guide the optimization toward improved approximation ratios. We evaluated this method on the Traveling Salesman Problem (TSP) using a QUBO-to-MaxCut transformation for up to $5$ layers. Our results show that Warm-PCE consistently outperforms standard PCE, achieving the optimum solution in $28\text{--}64\%$ of instances, versus $4\text{--}26\%$ for PCE, and attaining higher mean approximation ratios that improve with circuit depth. These findings highlight the practical value of this warm-start strategy for enhancing PCE-based solvers on near-term hardware.
 
\end{abstract}

\maketitle
\section{Introduction}
%
%
%
%

Combinatorial optimization is an area of vital importance for industry and technology \cite{papadimitriou1998combinatorial}. Problems such as vehicle routing \cite{BRAEKERS2016300}, supply chain planning \cite{SINGH20221636}, and network design \cite{Johnson1978TheCO} can be formulated as combinatorial optimization tasks, whose solutions directly translate into cost reductions, efficiency improvements, and competitive advantages \cite{korte2018combinatorial}. However, many of these problems are NP-Hard \cite{Karp1972}, so their exact solutions require computational resources that scale exponentially with the size of the problem, which challenge the classical approaches. 

Within this scenario, quantum optimization has emerged as a promising approach to tackle such problems, with the potential to deliver computational speedups or improved approximation ratios over classical methods \cite{Abbas2024}. In recent years, many methods have been proposed, particularly variational quantum algorithms (VQA) \cite{Cerezo_2021}, which are hybrid approaches that aim to combine the best of both worlds, classical and quantum, while addressing the limitations of current NISQ-era hardware \cite{preskill2018quantum}. These methods leave a set of free parameters to be optimized by classical algorithms, which can often cope with noise more effectively than their fully quantum counterparts. Among them, the Quantum Approximate Optimization Algorithm (QAOA) \cite{farhi2014quantum} and the Variational Quantum Eigensolver (VQE) \cite{peruzzo2014variational, Tilly_2022}, along with their various ansätze and extensions \cite{Wang2020XY, egger2021warm, tate2023warm, truger2024warm, Hadfield_2019}, are arguably the most studied ones. Despite their promise, the scalability of most VQAs is hampered by the widespread use of one-hot encoding schemes, which map each binary variable to a separate qubit \cite{Lucas_2014}. Although this approach simplifies the problem formulation, it results in prohibitively large qubit requirements for many relevant applications, severely constraining scalability on near-term hardware. This limitation motivates the search for more qubit-efficient encoding methods.

To address these limitations, more efficient encoding schemes have been proposed that map multiple binary variables to fewer qubits \cite{patti2022variational, sciorilli2025towards, Tan_2021, sciorilli2025competitivenisqqubitefficientsolver, sharma2025comparativestudyquantumoptimization, Huber_2024, leonidas2023qubitefficientquantumalgorithms}. One particularly promising method is Pauli Correlation Encoding \cite{sciorilli2025towards}, a variational solver for MaxCut problem \cite{korte2018combinatorial}, which achieves a polynomial reduction in qubit requirements relative to one-hot encoding by mapping classical variables in optimization problems to multi-body Pauli-matrix correlations. Furthermore, it is analytically demonstrated that PCE inherently mitigates barren plateaus \cite{sciorilli2025towards}, offering super-polynomial resilience against this phenomenon in training landscapes. Empirically, PCE has demonstrated competitive performance with state-of-the-art methods, making it an attractive candidate for near-term quantum optimization.

In this work, we extend the PCE framework by incorporating a warm-start strategy based on the Goemans--Williamson randomized rounding technique, akin to that proposed in~\cite{egger2021warm}. Our Warm-PCE variant introduces a bias in the cost function, guiding the optimization toward solutions with improved approximation ratios. We validate this approach on the Traveling Salesman Problem, a canonical NP-hard problem with broad industrial relevance in logistics and route optimization~\cite{korte2018combinatorial}. As this problem can be formulated as a quadratic unconstrained binary optimization (QUBO) problem~\cite{Lucas_2014}, we employ a standard QUBO-to-MaxCut transformation~\cite{barahona1989experiments, carmo2025warmstartingqaoaxymixers} and evaluate both PCE and Warm-PCE over circuit depths $p=1\text{--}5$. Across all depths, Warm-PCE consistently outperforms standard PCE: its per-graph success rate ranges from $28\%$ to $64\%$, whereas PCE achieves $4\%$--$26\%$ over the same range. A per-instance best-of-10 comparison further shows Warm-PCE winning the majority of graphs at every depth, with few ties. In terms of approximation quality, the mean approximation ratio of Warm-PCE increases monotonically with $p$ and, from $p \ge 3$, exceeds the PCE baseline by a clear margin. In contrast, PCE remains nearly flat across depths. These trends reinforce the benefit of warm-starting within PCE and motivate further extensions of the PCE-based framework.

To illustrate our results, this paper is organized as follows. Section~\ref{sec:pce} reviews Pauli Correlation Encoding and briefly explains how it can be applied to generic QUBO problems via a standard QUBO-to-MaxCut reduction. Section~\ref{sec:warm_pce} introduces Warm-PCE, our Goemans–Williamson–guided extension, detailing the modified objective and workflow. Section~\ref{sec:tsp} presents the Traveling Salesman Problem and instance generation. Section~\ref{sec:numerical_results} describes the experimental setup and evaluation metrics, and reports benchmarks on 5-city TSP instances under noise-free simulations. We then interpret the results, highlight Warm-PCE’s gains over PCE, and discuss applicability and limitations. Finally, Section~\ref{sec:conclusion} concludes and outlines directions for future work.

\section{Pauli Correlation Encoding}
\label{sec:pce}

Pauli correlation encoding is a framework recently introduced in \cite{sciorilli2025towards} where the main idea is to use the sign of Pauli strings like $\Pi_i$ of the form $X^{\otimes k} \otimes \mathbb{1}^{\otimes (n-k)}, \quad Y^{\otimes k} \otimes \mathbb{1}^{\otimes (n-k)}, \quad \text{or} \quad Z^{\otimes k} \otimes \mathbb{1}^{\otimes (n-k)}$ and its permutations, where $k$ is the number of correlations of Pauli strings and $n$ is the number of qubits, to encode each one of the binary variables $x_i$ from a combinatorial problem, that is,
\begin{equation}
    x_i := \mathrm{sgn}\!\left( \langle \Pi_i \rangle \right), \qquad i \in [m],
    \label{eq_sgn_pce}
\end{equation}
where $\mathrm{sgn}$ is the sign function and $\langle \Pi_i \rangle := \langle \Psi | \Pi_i | \Psi \rangle$ is the expectation value of $\Pi_i$ on a quantum state $|\Psi\rangle$. Here, $m$ denotes the number of encoded binary variables, which in principle can be as large as the number of independent Pauli correlations for $n$ qubits, i.e., $m \leq 4^n - 1$. So, instead of using one-hot encoding, where each binary variable is represented by one qubit in the $Z$ basis, which needs $n$ qubits for $N$ binary variables, using this approach we can compress the encoding achieving up to $3 \binom{n}{k}$ binary variables per qubit. For $k=2$, the case studied here, we have $m=\frac{1}{2}n(n-1)(n-2)$.

Originally, PCE focused on the weighted Maxcut problem. This is one of the most studied NP-Hard combinatorial problems \cite{Abbas2024}, especially for benchmarking proposes, whose goal is to divide the $m$ vertices from an undirected graph $G=(V,E)$ into two disjoint subsets in a way that maximizes the "cut", that is, the sum of edge weights $W_{ij}$ shared by the two subsets. \cite{papadimitriou1998combinatorial}  The problem can be mathematically formulated as the binary optimization,
\begin{equation}
    \max_{x \in \{-1,1\}^m} \; \sum_{i,j \in [m]} W_{ij} \big(1 - x_i x_j \big).
\end{equation}
where $x_i \in \{-1,1\}$ is the binary variable assigned to each vertex $i$ and $W_{ij}$ is the weight of each edge. This formulation is equivalent (up to an additive constant) to minimizing the quadratic form $x^\top W x$, i.e., a QUBO formulation \cite{papadimitriou1998combinatorial, Abbas2024}. 

As in all VQAs, we parametrize Eq.~\ref{eq_sgn_pce} as the output of a quantum
circuit with parameters~$\theta$, i.e., $\ket{\Psi}=\ket{\Psi(\theta)}$
\,(originally instantiated with a \emph{brickwork} architecture in the PCE formulation).
The parameter values are then optimized classically and updated in the circuit, yielding an interactive hybrid optimization process. However, instead of using the sign of the Pauli string as a binary variable, the cost function is rewritten in terms of smooth hyperbolic tangents to increase the learnability:
\begin{equation}
    \mathcal{L} = \sum_{(i,j) \in E} W_{ij} \, \tanh\!\big(\alpha \langle \Pi_i \rangle\big) 
\tanh\!\big(\alpha \langle \Pi_j \rangle\big) + \mathcal{L}^{(\mathrm{reg})},
\label{eq_pce}
\end{equation}
where the term $\mathcal{L}^{(\mathrm{reg})}$ is a regularization one that penalizes large correlator values, thereby constraining the optimizer to remain within the correlator domain where all possible bit-string solutions are representable, which in turn enhances the solver’s performance.

It is worth noting, particularly for the purposes of this paper, that although PCE was originally introduced as a MaxCut solver, it is a well-established fact that any QUBO problem can be transformed into a MaxCut instance. So, the potential applications can be much more general, encompassing a broad class of combinatorial optimization problems formulated as QUBOs. Moreover, PCE has recently been extended to general QUBO problems \cite{sharma2025comparativestudyquantumoptimization} and, in particular, to the Low Autocorrelation Binary Sequences problem \cite{sciorilli2025competitivenisqqubitefficientsolver}.

\section{Warm-PCE}
\label{sec:warm_pce}

Our proposal consists in extending the PCE’s cost function in Eq.~\ref{eq_pce} by
adding a bias term derived from the solution given by the Goemans--Williamson algorithm \cite{goemanswilliamson, karloffhoward}, steering the optimization towards solutions close to the GW one.
In the spirit of warm-start strategies (see, e.g., Egger et al.~\cite{egger2021warm}),
we temper this bias so that the search can deviate whenever it leads to better
optima: we introduce a small regularization hyperparameter $\varepsilon$ that
controls the strength of this bias. Formally, let
\begin{equation}
    s_i(\theta)=\tanh\!\big(\alpha \langle \Pi_i \rangle\big),
\end{equation}
and let $c_i^\star\in[0,1]$ denote the (possibly relaxed) GW bit for variable $i$.
To avoid over-biasing and numerical saturation, we regularize the GW bits with
$\varepsilon\in(0,0.5)$ as
\begin{equation}
    \widehat{c}_i^\star \;=\;
    \begin{cases}
        \varepsilon, & c_i^\star < \varepsilon,\\[2pt]
        \lfloor c_i^\star \rfloor, & \varepsilon \le c_i^\star \le 1-\varepsilon,\\[2pt]
        1-\varepsilon, & c_i^\star > 1-\varepsilon .
    \end{cases}
    \label{eq:gw_regularization}
\end{equation}
The resulting Warm-PCE objective is
\begin{equation}
    \begin{split}
        \mathcal{L}_{\mathrm{Warm\text{-}PCE}}(\theta)
        &= \sum_{(i,j)\in E}
           W_{ij}\,\Bigl(1+\bigl|\widehat{c}_i^\star-\widehat{c}_j^\star\bigr|\Bigr)\,
           s_i(\theta)\,s_j(\theta) \\
        &\quad + \mathcal{L}^{(\mathrm{reg})}.
    \end{split}
    \label{eq:warm_pce_loss}
\end{equation}

In short, we up-weight edges whose endpoints are assigned to different subsets by the GW solution. This biases the objective toward cuts that sever these edges, thereby increasing the cut value and guiding the optimizer toward solutions near the GW assignment. Note that setting $\varepsilon=0.5$ neutralizes the GW weight (it equals $1$ for all edges), recovering the standard PCE objective; smaller $\varepsilon$ increases the strength of the warm-start bias.

It's important to mention that, unlike the warm-start constructions of Egger et al.~\cite{egger2021warm}, which modify the initial state and the mixer so that the Goemans--Williamson rounding can be exactly recovered and the GW bound is preserved, our Warm-PCE injects GW information only as a soft bias in the objective function. Accordingly, we do not claim a formal guarantee that the GW cut will be recovered nor that the GW approximation ratio is preserved; the method is a heuristic that nudges the optimization toward the GW basin while allowing controlled deviations via~$\varepsilon$. We therefore view the proposed bias as a practical warm-start for PCE, whose benefits we substantiate empirically.

\begin{figure}[t]
  \includegraphics[width=\linewidth]{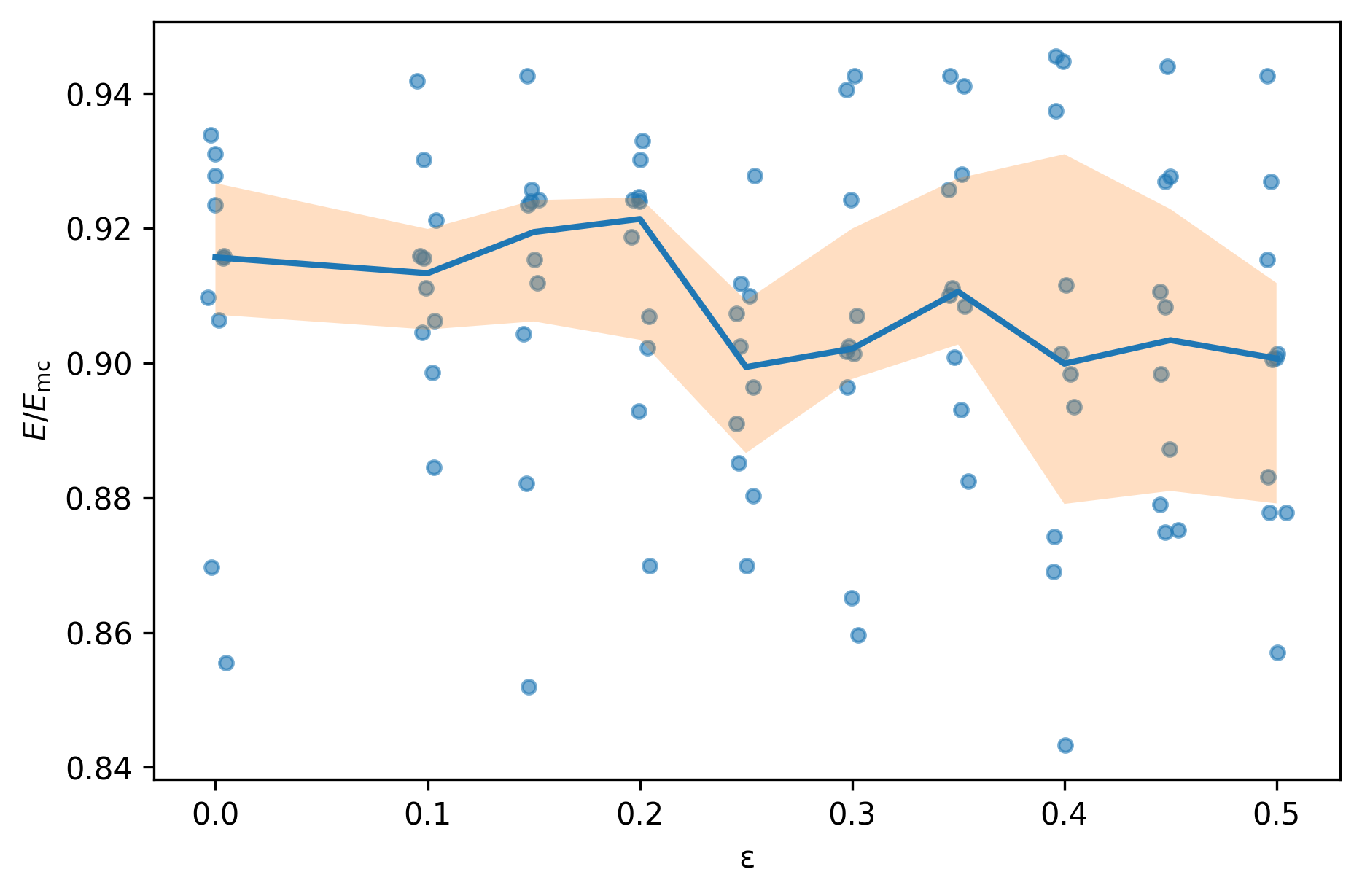}
    \caption{Sweep over the regularization parameter $\epsilon$. The y-axis shows the energy normalized by the maximum-cut energy ($E/E_{\mathrm{mc}}$). We evaluate 10 graphs; each dot corresponds to a single graph, aggregating 5 independent parameter initializations. The solid line is the median across graphs and the shaded band is the interquartile range (IQR). For the GW baseline, we perform 100 randomized roundings and keep the best cut (best-of-100). This results didn't use the classical bit-swap search as post-processing step.}
  \label{fig:epsilon_sweep}
\end{figure}

In order to understand the role of regularization parameter $\epsilon$ in this warm-starting approach for PCE and also select the most advantageous one, we performed a small sweep of $\varepsilon$ on a set of $10$ random $20$-vertex graphs, using $5$ independent initializations per instance (total of $50$ runs per value of $\varepsilon$). We show the results of this sweep in Figure~\ref{fig:epsilon_sweep} reports the ratio $E/E_{\mathrm{mc}}$, where $E$ is the objective obtained by Warm-PCE and $E_{\mathrm{mc}}$ is the best-known MaxCut value for the same instance computed using CPLEX. Each dot corresponds to a single run; the solid curve shows the median across runs, and the shaded band depicts the interquartile range (IQR). Performance shows a clear dependence on $\epsilon$, the median improves as $\epsilon$ increases up to $\epsilon \approx 0.2$ and then degrades for larger values, with variability widening notably for $\epsilon \gtrsim 0.35$. Note that $\epsilon = 0.5$ coincides with the standard PCE (the GW bias vanishes), which explains the drop in performance near this regime and supports the rationale behind Warm-PCE. Unless stated otherwise, we therefore adopt $\epsilon = 0.2$ as the default, which provides a moderate GW bias while remaining robust across instances and initializations. For the Goemans--Williamson baseline, we perform 100 independent randomized roundings and report the best cut among them (best-of-100).

Given the potential advantage of GW-biased warm starts for PCE, we next consider the Traveling Salesman Problem and briefly recall its QUBO formulation. Our objectives are twofold: (i) to quantify the advantage of Warm-PCE over standard PCE on small, practical instances, and (ii) to outline a qubit-efficient workflow that could, in principle, extend to larger graphs (more cities) and to other QUBO problems.

\section{Traveling Salesman Problem}\label{sec:tsp}

The Traveling Salesman Problem is a canonical NP-hard problem in combinatorial optimization~\cite{korte2018combinatorial}. Given a set of cities (nodes) and pairwise distances (edges), the objective is to determine the shortest Hamiltonian cycle, i.e., a tour that visits each city exactly once and returns to the starting point. In the symmetric case, the one we are considering in this work, the distance between any two cities is the same regardless of the traveling direction. Besides planning routes and logistics, TSP finds applications in many different areas such as DNA sequence and computer wiring \cite{CheikhrouhouKhoufi2021, NaleczCharkiewicz2022}.

\subsection{QUBO formulation}
We adopt the QUBO model of Lucas~\cite{Lucas_2014}. To remove symmetries and reduce the number of binary variables, we fix one city (say city $0$) as the starting and consequently the ending point of the tour. With this convention, the decision variable
\begin{equation}
x_{i,t}\in\{0,1\}
\end{equation}
indicates whether city $i$ ($i\in\{1,\dots,N-1\}$) is placed at position $t$ ($t\in\{1,\dots,N-1\}$) in the tour. This reduces the variable count from $N^2$ to $(N-1)^2$ \cite{Qian_2023}. For a $5$-city instance, which is the case studied here, the tour representation takes the form
\begin{equation}
    x = (1,2,3,4,5) \;\;\longrightarrow\;\; 1000 \, 0100 \, 0010 \, 0001,
\end{equation}
where the first city is fixed, leaving $4$ blocks of $4$ binary variables to encode the remaining cities. In the standard one-hot encoding, each one of those variables would be encoded in one qubit, totalizing $16$ qubits. Here, as we use the PCE scheme, this number will be much lower, as we will see.

The model enforces two assignment constraints to exclude infeasible solutions that do not correspond to valid tours:
\begin{align}
\sum_{t=1}^{N-1} x_{i,t} &= 1 &&\forall\, i\in\{1,\dots,N-1\}, \label{eq:tsp-city-once}\\
\sum_{i=1}^{N-1} x_{i,t} &= 1 &&\forall\, t\in\{1,\dots,N-1\}, \label{eq:tsp-one-per-slot}
\end{align}
where the first constraint ensures that each city is visited exactly once, and the second ensures that each position in the tour is occupied by exactly one city.

As usual in QUBO formulations, constraints are incorporated via quadratic penalties:
\begin{align}
H_{\mathrm{row}} &= \sum_{i=1}^{N-1}\left(\sum_{t=1}^{N-1} x_{i,t}-1\right)^{2}, \label{eq:Hrow}\\
H_{\mathrm{col}} &= \sum_{t=1}^{N-1}\left(\sum_{i=1}^{N-1} x_{i,t}-1\right)^{2}. \label{eq:Hcol}
\end{align}
and use a cost term that accounts for traveling between consecutive positions and the links to the fixed city $0$:
\begin{equation}
H_{\mathrm{cost}}
= \sum_{i,j=1}^{N-1}\sum_{t=1}^{N-2} W_{ij}\, x_{i,t}\,x_{j,t+1}
\;+\; \sum_{i=1}^{N-1} W_{0i}\,\big(x_{i,1}+x_{i,N-1}\big).
\end{equation}
The full QUBO Hamiltonian is
\begin{equation}
H_{\mathrm{QUBO}} \;=\; H_{\mathrm{cost}} \;+\; A\,H_{\mathrm{row}} \;+\; B\,H_{\mathrm{col}},
\end{equation}
with positive penalties $A,B$ chosen large enough to discourage violations of~\eqref{eq:tsp-city-once}–\eqref{eq:tsp-one-per-slot}.

\subsection{PCE Encoding}

Here, we follow the PCE encoding for $k=2$ correlations, which corresponds to a quadratic compression of the binary variables. Explicitly, for our case with $17$ binary variables, $16$ for the TSP encoding and $1$ more for the auxiliary node used in the QUBO-to-MaxCut conversion, we have:
\begin{align*}
x_0  &= Z_1 \otimes Z_2 \otimes \mathbb{1}_3 \otimes \mathbb{1}_4,\\
x_1  &= X_1 \otimes X_2 \otimes \mathbb{1}_3 \otimes \mathbb{1}_4,\\
x_2  &= Y_1 \otimes Y_2 \otimes \mathbb{1}_3 \otimes \mathbb{1}_4,\\
x_3  &= Z_1 \otimes \mathbb{1}_2 \otimes Z_3 \otimes \mathbb{1}_4,\\
x_4  &= X_1 \otimes \mathbb{1}_2 \otimes X_3 \otimes \mathbb{1}_4,\\
x_5  &= Y_1 \otimes \mathbb{1}_2 \otimes Y_3 \otimes \mathbb{1}_4,\\
x_6  &= Z_1 \otimes \mathbb{1}_2 \otimes \mathbb{1}_3 \otimes Z_4,\\
x_7  &= X_1 \otimes \mathbb{1}_2 \otimes \mathbb{1}_3 \otimes X_4,\\
x_8  &= Y_1 \otimes \mathbb{1}_2 \otimes \mathbb{1}_3 \otimes Y_4,\\
x_9  &= \mathbb{1}_1 \otimes Z_2 \otimes Z_3 \otimes \mathbb{1}_4,\\
x_{10} &= \mathbb{1}_1 \otimes X_2 \otimes X_3 \otimes \mathbb{1}_4,\\
x_{11} &= \mathbb{1}_1 \otimes Y_2 \otimes Y_3 \otimes \mathbb{1}_4,\\
x_{12} &= \mathbb{1}_1 \otimes Z_2 \otimes \mathbb{1}_3 \otimes Z_4,\\
x_{13} &= \mathbb{1}_1 \otimes X_2 \otimes \mathbb{1}_3 \otimes X_4,\\
x_{14} &= \mathbb{1}_1 \otimes Y_2 \otimes \mathbb{1}_3 \otimes Y_4,\\
x_{15} &= \mathbb{1}_1 \otimes \mathbb{1}_2 \otimes Z_3 \otimes Z_4,\\
x_{16} &= \mathbb{1}_1 \otimes \mathbb{1}_2 \otimes X_3 \otimes X_4.
\end{align*}

It is worth noting that there are other choices we can make to do this kind of encoding. One possibility could be to use only one Pauli matrix, $Z$, for instance, and just vary the number of correlations $k$. 

\section{Numerical Results and Analysis}\label{sec:numerical_results}

To compare the performance of our model with the standard PCE, we have performed numerical simulations with $50$ random graphs with $10$ initializations in each one, for layers $p=1$ to $5$. The classical optimizer used was COBYLA, with stopping criteria for up to $1000$ interactions. Regarding the GW results for the classical bias, we have considered the best out of $100$ randomized roundings as before. Also, here we choose to use a TwoLocal circuit ansatz for both PCE and Warm-PCE, instead of the original brickwork design. For this comparison we keep the original full PCE pipeline, using the classical bit-swap search as post-processing step \cite{sciorilli2025towards}.

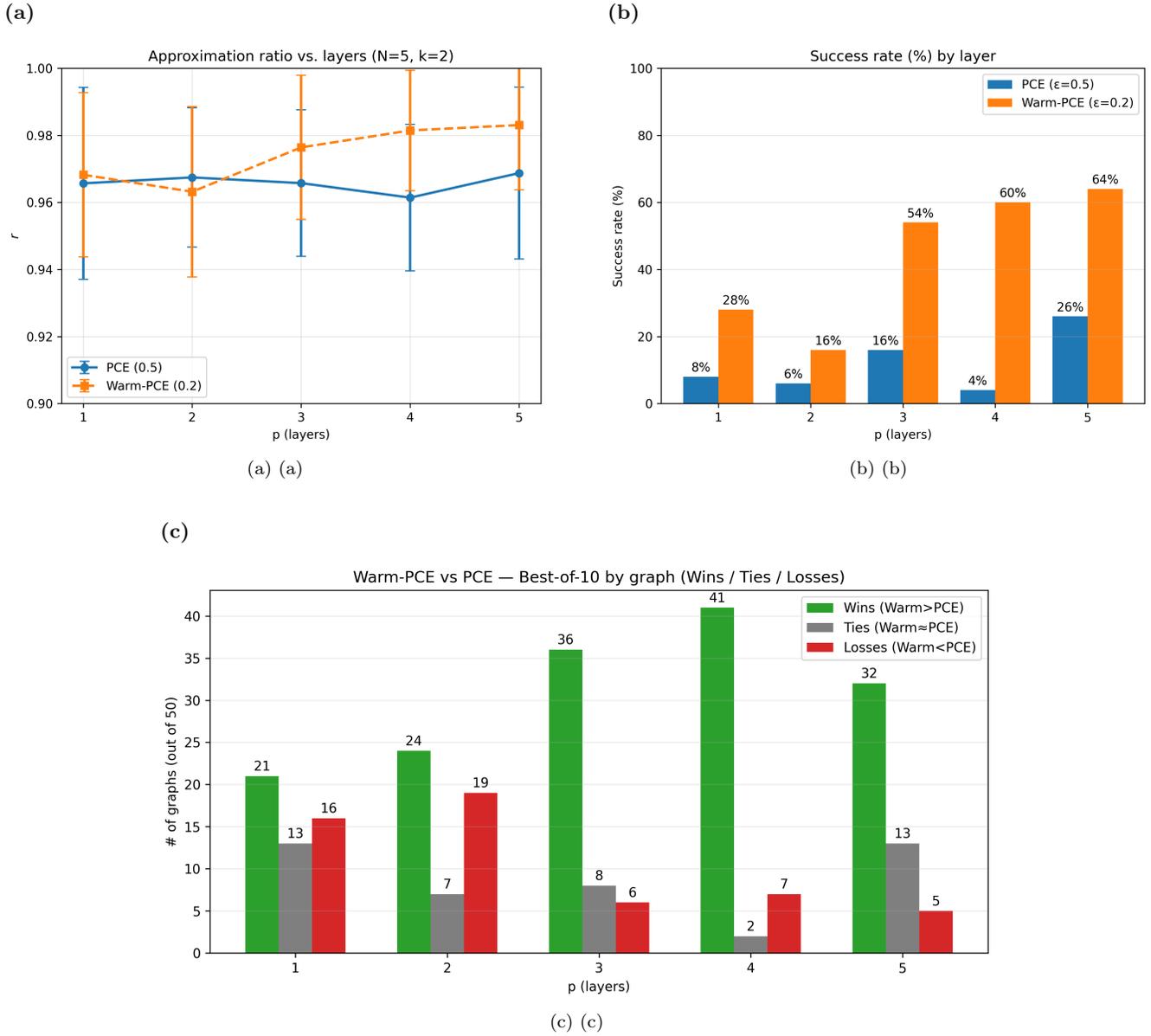
\begin{figure*}[!t]
  \centering

  \subfloat[(a)\label{fig:ratio_p}]{
    \begin{tikzpicture}
      \node[inner sep=0] (img)
        {\includegraphics[width=0.47\linewidth]{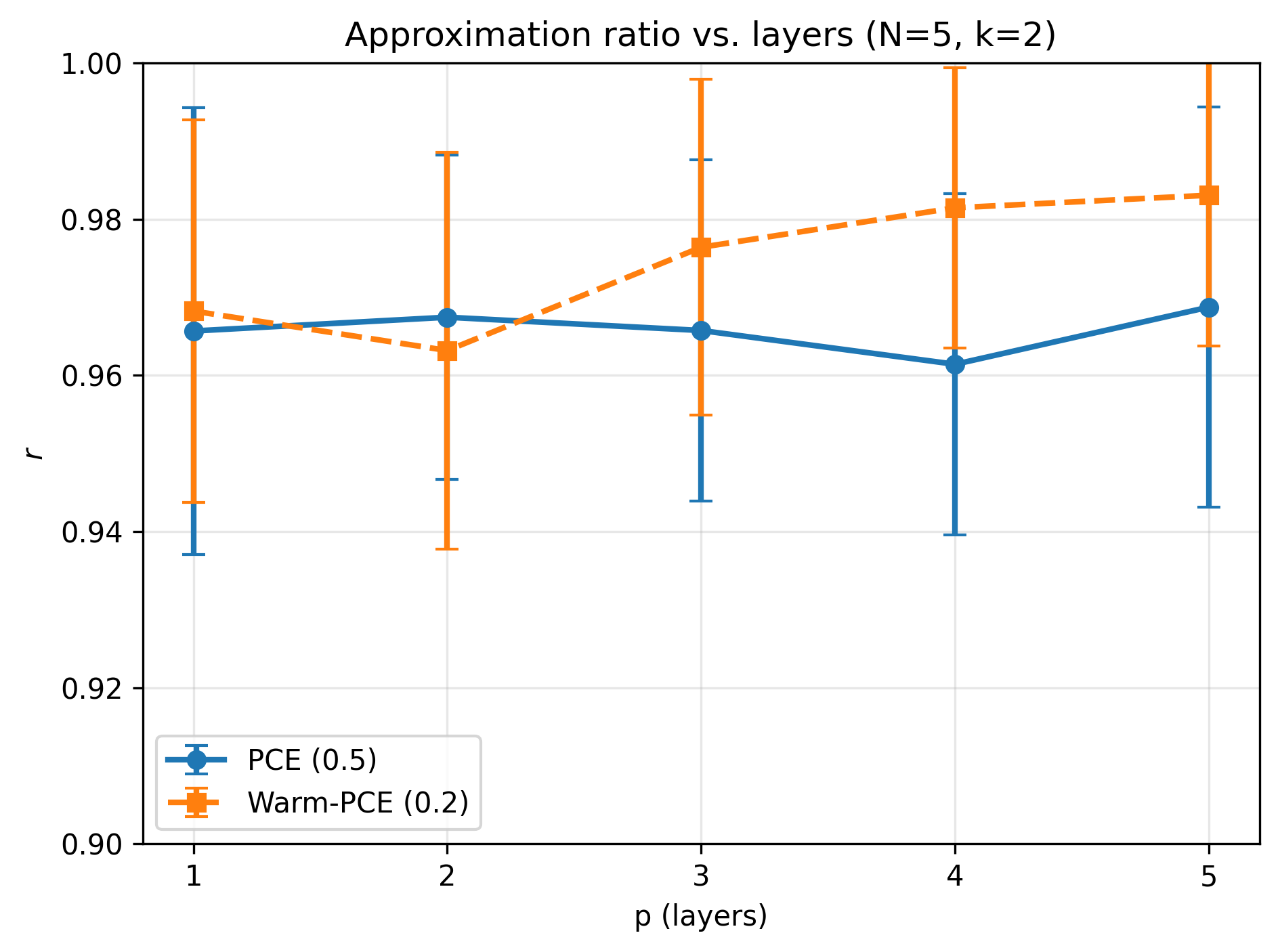}};
      \node[
        anchor=south west, xshift=0pt, yshift=6pt,
        font=\bfseries\small, fill=white, rounded corners=1pt,
        inner sep=1.5pt, fill opacity=0.95
      ] at (img.north west) {(a)};
    \end{tikzpicture}
  }\hfill
  \subfloat[(b)\label{fig:success_p}]{
    \begin{tikzpicture}
      \node[inner sep=0] (img)
        {\includegraphics[width=0.47\linewidth]{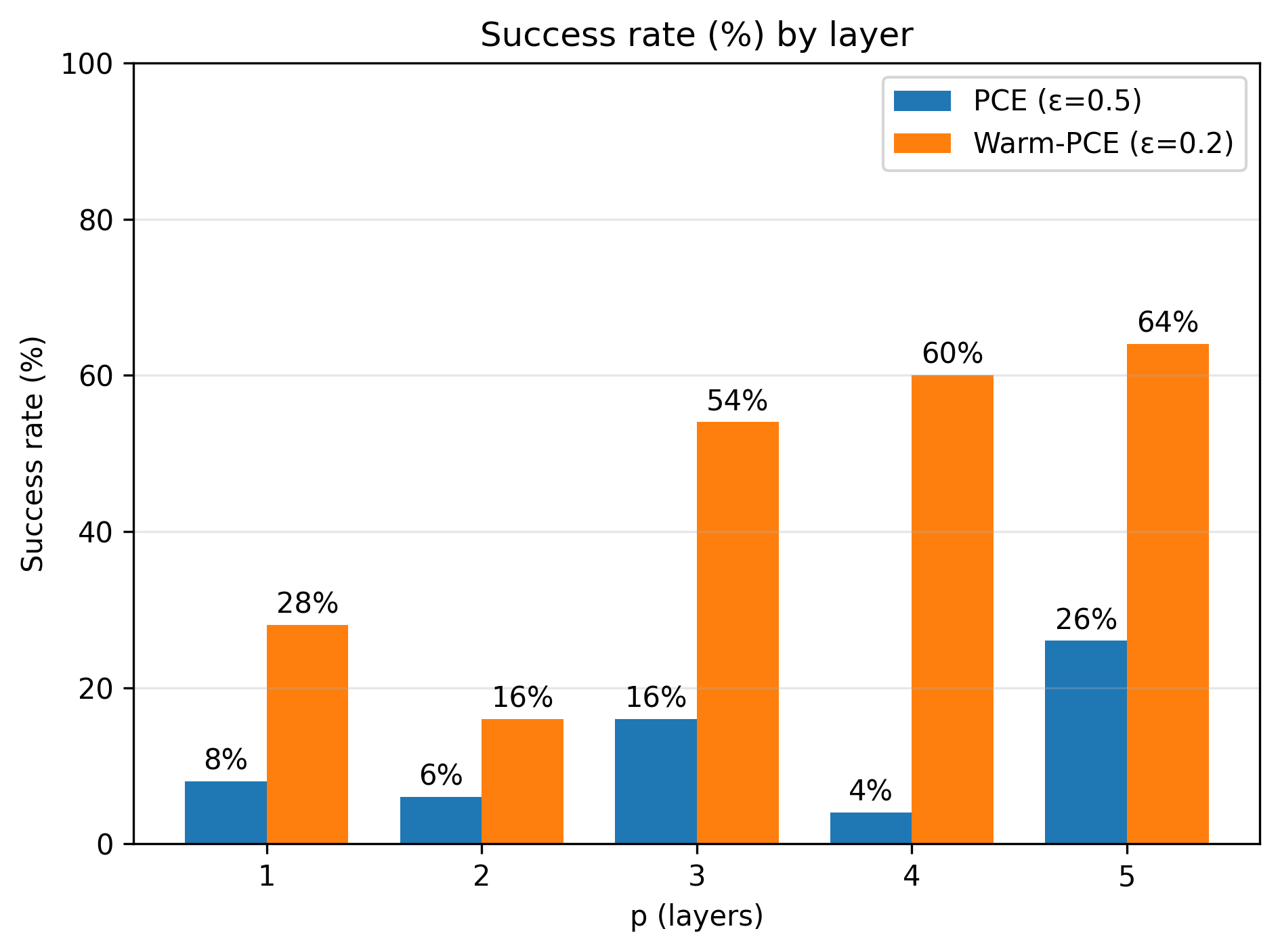}};
      \node[
        anchor=south west, xshift=0pt, yshift=6pt,
        font=\bfseries\small, fill=white, rounded corners=1pt,
        inner sep=1.5pt, fill opacity=0.95
      ] at (img.north west) {(b)};
    \end{tikzpicture}
  }

  \par\medskip

  \subfloat[(c)\label{fig:wins_ties_loss}]{
    \begin{tikzpicture}
      \node[inner sep=0] (img)
        {\includegraphics[width=0.72\linewidth]{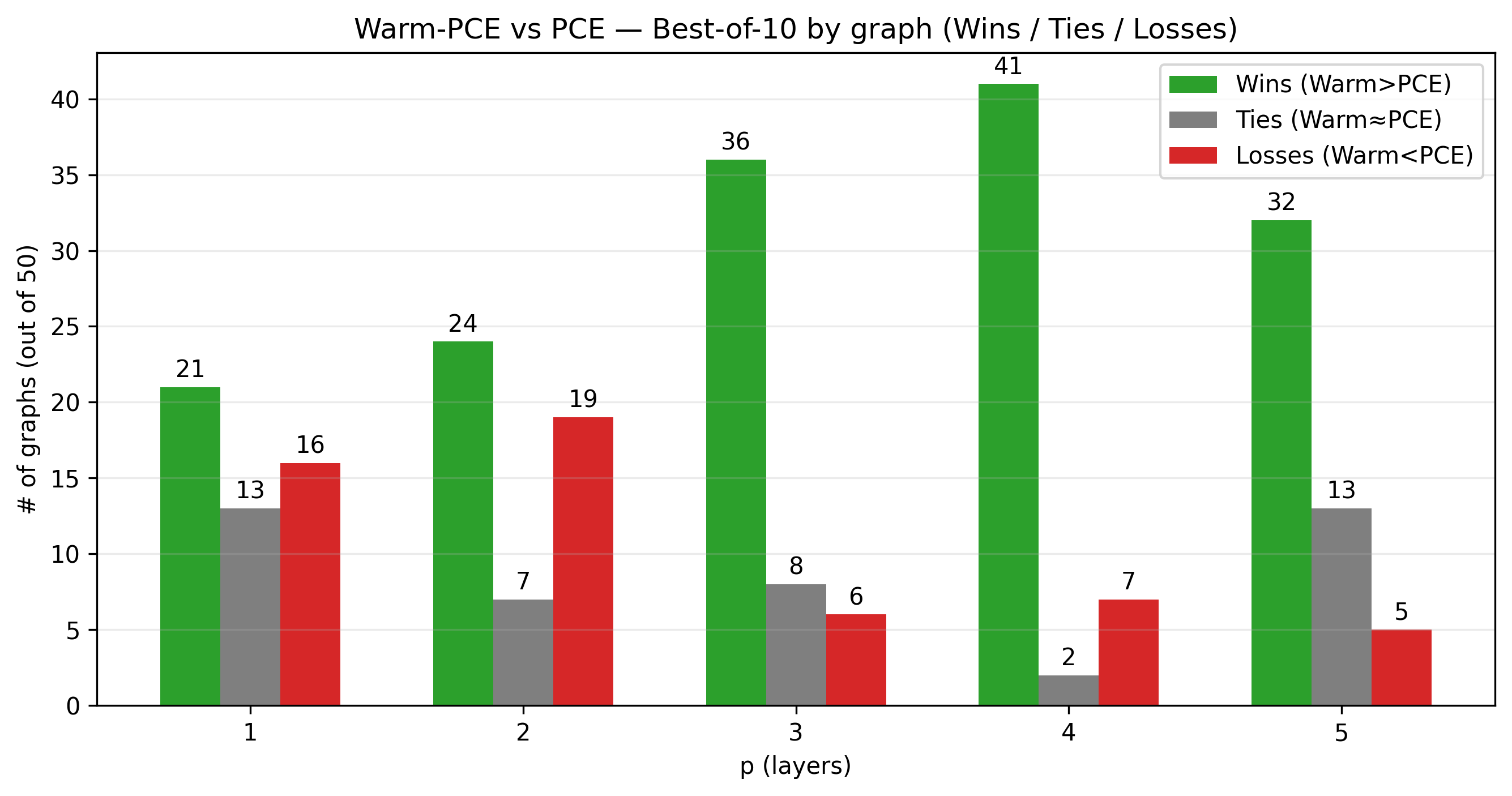}};
      \node[
        anchor=south west, xshift=0pt, yshift=6pt,
        font=\bfseries\small, fill=white, rounded corners=1pt,
        inner sep=1.5pt, fill opacity=0.95
      ] at (img.north west) {(c)};
    \end{tikzpicture}
  }

  \caption{Warm–PCE vs.\ PCE (5-city TSP).
  (a) Mean approximation ratio $r$ vs.\ depth $p$ (Warm–PCE improves monotonically with depth; PCE is nearly flat).
  (b) Success rate of hitting the optimum at least once across 10 initializations (28–64\% vs.\ 4–26\%; $\sim\!2\times$–$15\times$ gain; standout at $p{=}4$: 60\% vs.\ 4\%).
  (c) Best-of-10 per graph: Warm–PCE wins on all depths, with emphasis on 36–41/50 graphs for $p{=}3$–$4$ and 32/50 for $p{=}5$ (13 ties). All these results consist in valid routes for TSP.}
  \label{fig:all_results}
\end{figure*}

The numerical experiments reveal a consistent and growing advantage of Warm-PCE over the standard PCE as the circuit depth increases. In terms of approximation quality, Warm-PCE shows a clear monotonic improvement with $p$ beyond the third layer, while PCE remains nearly flat across depths. This widening gap (Fig.~\ref{fig:ratio_p}) indicates that GW-based bias not only improves average performance, but also leverages circuit depth more effectively than the unbiased formulation. Considering that barren plateaus are typically expected to emerge only at depths around $8.5n$ for $n$ qubits, as reported in \cite{sciorilli2025towards}, there remains substantial room for further gains as depth increases. Although it is necessary to investigate the behavior of Warm-PCE in deep layers, clever initializations have been reported as a potential but still unknown way to explore an exponentially large space \cite{Cerezo_2025}, i.e. without falling into polynomial subspaces, which will imply classical simulability.

\FloatBarrier

Beyond average ratios, the contrast is striking when we consider the probability of hitting the true optimum at least once over ten initializations. Warm–PCE achieves success rates of $28\text{–}64\%$ versus $4\text{–}26\%$ for PCE, which means a $\sim2\times$–$15\times$ increase depending on depth (Fig.~\ref{fig:success_p}). The advantage is already substantial for $p\ge3$ $(54\text{–}64\%$ vs.\ $16\text{–}26\%$) and becomes dramatic at $p=4$ $(60\%$ vs.\ $4\%$), underscoring the practical superiority of Warm–PCE over standard PCE under the same qubit budget.

A per-instance analysis further confirms the robustness of the approach. Considering each graph individually and focusing in the best result per instance, even though it is not the optimum path, Warm-PCE secures substantially more wins than losses across all depths, with ties occurring only marginally (Fig.~\ref{fig:wins_ties_loss}). The dominance is especially clear for intermediate depths ($p=3,4$), where Warm-PCE outperforms the baseline on the vast majority of instances, winning on $36$–$41$ out of $50$ graphs. Taken together, these complementary perspectives highlight not only higher average quality but also greater reliability of Warm-PCE in recovering the best approximate solution, which for TSP instances we used was always equivalent to valid routes, reinforcing the potential of warm-start strategies within the PCE framework.

\section{Conclusion}
\label{sec:conclusion}


In this study, we introduced a enhanced version of the PCE, using a classical bias from Goemans-Williamson randomized rounding into the algorithm’s loss function, which we called Warm-PCE. On 5-city TSP instances, our results comparing it with the standard PCE have shown a clear advantage of our method from at least three complementary perspectives: (i) the mean approximation ratio increases monotonically with depth for $p\ge 3$ while PCE remains nearly flat, yielding higher ratios overall; (ii) the probability of reaching the true optimum improves between $\sim2\times$–$15\times$ the true solutions achieved by PCE depending on depth; and (iii) in per-instance best-of-10 comparisons Warm-PCE wins on the vast majority of graphs, especially at $p\in\{3,4\}$, indicating that even when the global optimum is not reached, Warm-PCE returns better approximate solutions than PCE. All simulation code is openly available; see Ref. \cite{warm_pce_code}.

Although our experiments centered on small TSP instances, the proposed warm bias is problem-agnostic and applies to any QUBO (or directly to MaxCut), just as in the original PCE. Since plain PCE already attains performance competitive with leading classical heuristics such as Burer–Monteiro and follow-ups~\cite{Burer2003,Dunning2018}, a natural next step is a systematic evaluation of Warm-PCE on the standard large-scale Max-Cut benchmark suites, exploring deeper circuits and higher-order encodings ($k>2$). If the gains we observe at shallow depth persist, Warm-PCE could match, or even surpass, the original PCE while retaining its qubit savings. For TSP, it will be equally interesting to stress-test the method on substantially larger instances, quantifying the trade-offs between depth, $k$, and solution quality, and assessing robustness under noise. More broadly, warm biasing seems to be able to reduce the depth needed to attain high-quality solutions relative to standard PCE, easing sampling demands and facilitating hybrid quantum–classical workflows on current hardware. This underscores the practical value of warm starts and paves the way for further extensions and broader applications of the PCE-based framework toward larger benchmark graphs and large-scale combinatorial instances on next-generation processors.

\twocolumngrid

\appendix

\FloatBarrier

\bibliographystyle{apsrev4-2}
\bibliography{bibl}

\end{document}